\documentclass{aa}
\usepackage{graphicx,txfonts, tikz}
\usetikzlibrary{shapes,arrows,chains, calc,backgrounds,fit}

\begin{document}

\title{Degeneracies of parametric lens model families near folds and cusps}
\titlerunning{}
\author{J. Wagner \and M. Bartelmann}
\institute{Universit\"at Heidelberg, Zentrum f\"ur Astronomie, Institut f\"ur Theoretische Astrophysik, Philosophenweg 12, 69120 Heidelberg, Germany\\ \email{j.wagner@uni-heidelberg.de}}
\date{Received XX; accepted XX}

\abstract{We develop an approach to select families of lens models that can describe doubly and triply gravitationally lensed images near folds and cusps using the model-independent ratios of lensing-potential derivatives derived in \cite{bib:Wagner}. Models are selected by comparing these model-independent ratios of potential derivatives to (numerically determined) ratios of potential derivatives along critical curves for entire lens model families in a given range of parameter values. This comparison returns parameter ranges which lens model families can reproduce observation within, as well as sections of the critical curve where image sets of the observed type can appear. If the model-independent potential-derivative ratios inferred from the observation fall outside the range of these ratios derived for the lens model family, the entire family can be excluded as a feasible model in the given volume in parameter space. We employ this approach for the family of singular isothermal spheres with external shear to examples of lensing by a galaxy and two galaxy clusters (JVAS B1422+231, SDSS J2222+2745, and MACS J1149.5+2223) and show that the results obtained by our general method are in good agreement with results of previous model fits.}


\maketitle

%%%%%%%%%%%%%%%%%%%%%%%%%%%
\section{Introduction and motivation}
\label{sec:introduction}

In \cite{bib:Wagner} we derived the model-independent information contained in strongly-lensed configurations of point-like or extended images. We obtained (ratios of) derivatives of the lensing potential in the vicinity of fold and cusp singularities on critical curves. This approach, briefly summarized in Sect.~\ref{sec:basics} below, uses the relative separations between the lensed images, magnification ratios and ellipticities of extended images, and time delays between images of sources whose intensity is varying in time.

Based on these results, we now develop an approach to constrain volumes in parameter space for lens-model families able to reproduce observed image configurations. This approach also allows to locate sections on the critical curve where the images can be located. We discuss in Sect.~\ref{sec:model_selection} the lens model families which our approach can be applied to. In Sect.~\ref{sec:implementation}, we describe the implementation details of our algorithm to compare the model-independent potential-derivative ratios with those ratios for lens-model families, and determine the allowed parameter ranges and the possible positions of the image pair or triple along the critical curve. Examples of strong lensing by a galaxy (JVAS B1422+231) and two galaxy clusters (MACS J1149.5+2223 and SDSS J2222+2745) in Sect.~\ref{sec:examples} illustrate how our approach can find ranges of possible parameter values or exclude a family of lens models with given parameter ranges. We summarize and conclude in Sect.~\ref{sec:conclusion} and discuss how data processing methods could be adapted to make our model selection method more reliable.

%%%%%%%%%%%%%%%%%%%%%%%%%%%
\section{Model-independent potential-derivative ratios}
\label{sec:basics}

\subsection{Summary of earlier results}

Using the notation introduced by \cite{bib:SEF} or \cite{bib:Petters}, let $\phi(x,y)$ be the Fermat potential, $x, y \in \mathbb{R}^2$, of a sufficiently smooth gravitational-lens model and $(x^{(0)}, y^{(0)})$ a singular point, for which the Hessian of $\phi(x,y)$ has rank 1. Then, there are two types of stable singular points on a critical curve characterised by this Hessian: folds which have a non-vanishing tangent vector in the source plane and cusps for which this tangent vector vanishes.

Without loss of generality, we can choose coordinate systems in the image and source planes with their origins shifted to $(x^{(0)},y^{(0)})$ and rotated such that
\begin{equation}
 \phi_1^{(0)} = \phi_2^{(0)} = \phi_{12}^{(0)} = \phi_{22}^{(0)} = 0\;,
\label{eq:coordinate_system}
\end{equation}
where we abbreviate
\begin{equation}
  \left.\frac{\partial\phi}{\partial x_i}\right\vert_{(x^{(0)},y^{(0)})} =
  \phi_i^{(0)} \quad i = 1,2 \;.
\label{eq:abbrev}
\end{equation}
In such coordinate systems, the following relations hold at a fold singularity,
\begin{align}
  \phi_{222}^{(0)} &= \dfrac{12ct_\mathrm{d}^{(AB)}D_{\mathrm{ds}}}{D_\mathrm{d}D_\mathrm{s}  (1+z_\mathrm{d})(\delta_{AB2})^3}\;,
  \label{eq:time_delay_fold} \\
  \dfrac{\phi_{222}^{(0)} }{\phi_{11}^{(0)}} &= \dfrac{2r_A}{\delta_{AB2}} \label{eq:ratio_fold}
\end{align}
where $c$ denotes the speed of light, $z_\mathrm{d}$ the redshift of the lens plane, $t_\mathrm{d}^{(AB)}$ the measured time delay, $D_\mathrm{ds}$, $D_\mathrm{d}$, and $D_\mathrm{s}$ the angular diameter distances between the lens and the source plane, the observer and the lens, and the observer and the source, respectively \citep{bib:Wagner}. $\delta_{AB2} = x_{A2} - x_{B2}$ is the separation between (the centres of brightness of) the two images $A$ and $B$ at a fold in the lens plane, and $r_A$ denotes the ratio of the semi-major to the semi-minor axis of image $A$.

For images near a cusp singularity, we found
\begin{align}
  \phi_{2222}^{(0)} &= \dfrac{8ct_\mathrm{d}^{(ij)}D_{\mathrm{ds}}}{D_\mathrm{d} D_\mathrm{s} (1 + z_\mathrm{d}) (\delta_{ij2})^4} \quad (i, j = A, B, C, \; i \ne j)\;,
\label{eq:time_delay_cusp} \\
  \dfrac{\phi_{122}^{(0)}}{\phi_{11}^{(0)}} &= \dfrac{2\left( \delta_{AB1}\delta_{AC2} - \delta_{AC1}\delta_{AB2} \right)}{\delta_{AB2}\delta_{AC2}\left( \delta_{AB2} - \delta_{AC2} \right)}\;,
\label{eq:ratio1_cusp} \\
  \dfrac{\phi_{2222}^{(0)}}{\phi_{11}^{(0)}} &= \dfrac{2}{(\delta_{ij2})^2}\left( \dfrac{\phi_{122}^{(0)}}{\phi_{11}^{(0)}}\delta_{ij1} -r_i + r_j \right)\;;
\label{eq:ratio2_cusp}
\end{align}
where the distances $\delta_{ij1}, \delta_{ij2}$, $i,j=A, B, C$, are given by
\begin{align}
\delta_{AB1} =& -\delta_{AB} \, \cos\left(\dfrac{\alpha_A}{2}\right)\;, &\delta_{AB2} =& -\delta_{AB} \, \sin \left(\dfrac{\alpha_A}{2}\right) \;, \label{eq:cusp_trigonometry1}\\
\delta_{AC1} =& -\delta_{AC} \, \cos\left(\dfrac{\alpha_A}{2}\right)\;, &\delta_{AC2} =& \phantom{-} \delta_{AC} \, \sin \left(\dfrac{\alpha_A}{2}\right) \;, \\
\delta_{BC1} =& - \delta_{BC} \, \cos\left(\dfrac{\alpha_A + \alpha_B}{2}\right)\;, &\delta_{BC2} =& \delta_{BC} \, \sin\left(\alpha_B + \dfrac{\alpha_A}{2}\right)
\end{align}
and $\alpha_i$ denotes the observable angle at the vertex $i$ of the triangle spanned by $A$, $B$, and $C$.

%%%%%%%%%%
\subsection{Estimates of errors and uncertainties}
\label{sec:errors}

Evaluating Eqs.~\ref{eq:time_delay_fold}, \ref{eq:ratio_fold}, \ref{eq:time_delay_cusp}, \ref{eq:ratio1_cusp}, and \ref{eq:ratio2_cusp} for all possible combinations of observables yields an interval of potential-derivative ratios whose width indicates their uncertainties because we expect the results to be independent of the specific image pair they are derived from, as further detailed in \cite{bib:Wagner}.

%%%%%%%%%%%%%%%%%%%%%%%%%%%
\section{Model selection}
\label{sec:model_selection}

A family of lens models with $n$ parameters $p = (p_1, ..., p_n) \in \mathbb{R}$ is given by the Fermat potential
\begin{equation}
\phi(x,y,p) = \dfrac12 \left(x-y \right)^2 - \psi(x,p) \,
\label{eq:lensing_potential}
\end{equation}
with the deflection potential $\psi(x,p)$. Then, solving
\begin{equation}
\phi_{11}^{(0)}(p) \, \phi_{22}^{(0)}(p) - \left( \phi_{12}^{(0)}(p) \right)^2 = 0\;,
\label{eq:critical_curve}
\end{equation}
the singular points $x^{(0)}(p)$ of the lens mapping by the potential $\phi$ are found. The existence, type and number of critical curves formed by these singular points will generally depend on the choice of $p$, as further detailed in \cite{bib:SEF} and \cite{bib:Petters}. Since Eq.~\ref{eq:critical_curve} cannot be solved analytically for general $p$ and arbitrary lens models, specific models and physically reasonable parameter ranges are usually chosen to solve Eq.~\ref{eq:critical_curve} numerically.

Thus, for a given set of relative image positions and image ellipticities, the right-hand sides of Eqs.~\ref{eq:ratio_fold}, \ref{eq:ratio1_cusp} and \ref{eq:ratio2_cusp} can be evaluated, and their left-hand sides can be derived numerically for any lens-model family once ranges are specified for the parameters to vary in. These parameter ranges need to be adapted to the case at hand, e.g.\ lensing by a galaxy or a galaxy cluster, and they may be adapted to models previously constructed (see Sect.~\ref{sec:implementation_sise} for examples).

Eqs.~\ref{eq:ratio_fold}, \ref{eq:ratio1_cusp}, and \ref{eq:ratio2_cusp}indicate that only one parameter can be determined per image pair at a fold singularity, and two at a cusp. This is no weakness of our method, but a general limitation as long as no higher-order terms in the Taylor expansion of the lensing potential near critical points can reliably be measured. We thus restrict our analysis to lens-model families with two parameters. For illustration, we focus here on a singular isothermal sphere (SIS) with external shear (SIS+E). It can be considered a representative family of elliptical lens models since \cite{bib:Kovner} showed that axisymmetric lens models with external shear can often describe elliptical mass distributions equally well and the latter, in turn, can be related to elliptical potentials as introduced by \cite{bib:Kassiola}.

Compared to elliptical mass distributions or elliptical potentials, the computational effort for calculating critical curves or image positions is reduced because all lensing characteristics of the SIS can be analytically determined. Furthermore, adding external shear independent of the lens has been shown to result in more realistic lens models compared to axisymmetric or purely elliptical models, as demonstrated by \cite{bib:Keeton}.

%%%%%%%%
\subsection{Singular isothermal sphere with external shear}

Due to the axial symmetry of the lens (SIS), the external shear can be aligned with the coordinate axes of the lens without loss of generality. Transforming to plane polar coordinates $(r, \theta)$, the deflection potential $\psi$ of an SIS+E can then be expressed as
\begin{equation}
\psi(r, \theta, a, \Gamma_1) = a r - \dfrac12 \Gamma_1 r^2 \cos(2 \theta)\;, \quad a = 4\pi \left(\dfrac{\sigma}{c}\right)^2 \dfrac{D_{\mathrm{ds}}}{D_\mathrm{s}} \;,
\label{eq:lensing_potential_sise}
\end{equation}
where $\Gamma_1$ is the amplitude of the external shear, and $a$ is determined by the velocity dispersion along the line-of-sight $\sigma$ relative to $c$, the angular-diameter distance between the lens and the source $D_{\mathrm{ds}}$, and between the observer and the source $D_{\mathrm{s}}$.

The critical curves $r^{(0)}(\theta, a, \Gamma_1)$ are given by
\begin{equation}
r^{(0)}(\theta, a, \Gamma_1) = \dfrac{a \left( \Gamma_1 \cos(2 \theta) + 1 \right)}{1 - \Gamma_1^2} \;.
\label{eq:critical_curve_sise}
\end{equation}
In the coordinate system chosen, the cusp singularities are located at $\theta = 0, \pi/2, \pi, 3/2 \pi$, connected by fold segments given by Eq.~\ref{eq:critical_curve_sise} for the remaining values of $\theta$ between 0 and $2\pi$. Due to the elliptical symmetry of the potential, it suffices to consider the first quadrant, i.e.\ we restrict $\theta$ to $\left[ 0, \pi/2 \right]$.

Writing $r^{(0)}(\theta,a,\Gamma_1)$ in Cartesian coordinates and transforming into the local coordinate system where Eqs.~\ref{eq:coordinate_system} hold (see Sect.~\ref{sec:implementation} for the transformation matrix), $\phi_{222}^{(0)}/\phi_{11}^{(0)}$, $\phi_{122}^{(0)}/\phi_{11}^{(0)}$ and $\phi_{2222}^{(0)}/\phi_{11}^{(0)}$ can be determined for $\theta \in \left[ 0, \pi/2 \right]$ once $a$ and $\Gamma_1$ are set.

%%%%%%%%%%%%%%%%%%%%%%%%%%%
\section{Implementation details}
\label{sec:implementation}

\subsection{Calculating the model-dependent potential-derivative ratios}
\label{sec:database}

Although we focus on the two-parameter family of singular isothermal lens models with external shear, the subsequent procedure to derive the potential-derivative ratios required for the left-hand sides of Eqs.~\ref{eq:ratio_fold}, \ref{eq:ratio1_cusp} and \ref{eq:ratio2_cusp} can be applied to arbitrary lens models as defined in Eq.~\ref{eq:lensing_potential}. Fig.~\ref{fig:flowchart} summarises the individual steps for an arbitrary lensing potential $\phi(x,y,p)$ with an unspecified number of parameters $p = (p_1, ..., p_n)$.

\begin{enumerate}
% 1.
\item A family of lens models $\phi(x,y,p)$ is chosen that seems likely to produce the given set of observed images, e.g.\ using Eq.~\ref{eq:lensing_potential_sise}.
% 2.
\item Physically reasonable values for all parameters are identified based on the type of lens as outlined in Sect.~\ref{sec:model_selection}. For instance, these values can be sampled from a range of values or taken from previous model fits. Thus, all following steps are performed for a given set of fixed parameter values $p_0$ (see Sect.~\ref{sec:implementation_sise} for the case of SIS+E). 
% 3.
\item The critical curves of the lens mapping are found, and their type (fold or cusp) is determined depending on the given parameter values $p_0$. For axially or elliptically symmetric lens models, a conversion to polar coordinates is advantageous for solving the equation for the critical radius $r^{(0)}(\theta, p_0)$. In this case, the equation to be solved for the singular points $(r^{(0)}, \theta^{(0)})$ reads
\begin{equation}
1 - \psi_{rr} - \dfrac{\psi_{r}}{r} - \dfrac{\psi_{pp}}{r^2} +\psi_{rr} \left( \dfrac{\psi_r}{r}+\dfrac{\psi_{pp}}{r^2}\right) - \dfrac{1}{r^2} \left( \dfrac{\psi_p}{r} - \psi_{rp} \right)^2 = 0\;,
\label{eq:general_detA}
\end{equation}
where the indices denote the derivatives of the deflection potential (that still depends on $p_0$) in the direction of the indexed variable. The points $(r^{(0)}(p_0), \theta^{(0)}(p_0))$ found are transformed to Cartesian coordinates for the next steps.
% 4.
\item Having obtained the fold and cusp points, we transform to the coordinate system required for Eq.~\ref{eq:coordinate_system} to hold, i.e.\ where the distortion matrix containing the second-order potential derivatives is diagonal and the origin is placed on the critical curve. To do so, all second, third and for the cusp also fourth order derivatives of $\phi(x,y,p_0)$ are determined. Since the cusps have already been located on the coordinate axes in the coordinate system of the isothermal lens model with external shear by construction,
\begin{align}
& \left( \phi_{11}^{(0)}(p_0) \ne 0 \quad \wedge \quad \phi_{12}^{(0)}(p_0) = \phi_{22}^{(0)}(p_0) = 0\right) \quad \vee \label{eq:system_translated}\\
&\left( \phi_{22}^{(0)}(p_0) \ne 0 \quad \wedge \quad \phi_{11}^{(0)}(p_0) =\phi_{12}^{(0)}(p_0) = 0 \right) \label{eq:system_rotated}
\end{align}
hold. This implies that the coordinate system of Eq.~\ref{eq:coordinate_system} can be attained by translating the singular point to the origin (Eq.~\ref{eq:system_translated}), or can be attained by a counter-clockwise rotation by $\pi/2$ followed by the translation to the origin (Eq.~\ref{eq:system_rotated}). To rotate a fold point not located on the axis (i.e.\ $\phi_{12}^{(0)}(p_0) \ne 0$) into this coordinate system, a rotation by the matrix
\begin{equation}
R = \left( \begin{matrix} \alpha & - \beta \\ \beta & \alpha \end{matrix}\right)
\end{equation}
with
\begin{equation}
\alpha = \dfrac{\phi_{11}^{(0)}(p_0)}{\phi_{12}^{(0)}(p_0)} \beta \;, \quad \beta = \sqrt{\dfrac{\phi_{22}^{(0)}(p_0)}{\phi_{11}^{(0)}(p_0) + \phi_{22}^{(0)}(p_0)}}
\label{eq:alpha_beta}
\end{equation}
needs to be applied. Should a fold point lie on one of the axes, the same procedure as for the cusps can be applied. Due to the elliptical symmetry of the SIS+E lens model, the rotation matrix is uniquely determined since $R$ has to be orthonormal with a rotation angle less than $\pi$.
% 5.
\item Given the rotation matrix $R$, the potential derivatives of interest have to be transformed into the coordinate system of Eq.~\ref{eq:coordinate_system}. The detailed transformation equations are shown in the appendix. After this final transformation, the potential-derivative ratios as required for the left-hand sides of Eqs.~\ref{eq:ratio_fold}, \ref{eq:ratio1_cusp} and \ref{eq:ratio2_cusp} are determined.
% 6.
\item  For each lens model with specific parameter values $p_0$, and for each fold and cusp point on the critical curve, the ratios of potential derivatives are saved in a database, and steps 2 to 6 are repeated for each lens model until the potential derivatives of each critical point for all previously selected parameter values are determined.
\end{enumerate}

% -------------------------------------------------
% Set up a new layer for the debugging marks, and make sure it is on
% top
\pgfdeclarelayer{background}
\pgfsetlayers{background,main}
% -------------------------------------------------
% Start the picture
\begin{figure*}
\begin{center}
\begin{tikzpicture}[%
    >=triangle 60,              % Nice arrows; your taste may be different
    start chain=going below,    % General flow is top-to-bottom
    node distance=6mm and 60mm, % Global setup of box spacing
    every join/.style={norm},   % Default linetype for connecting boxes
    ]
% -------------------------------------------------
% A few box styles
% <on chain> *and* <on grid> reduce the need for manual relative
% positioning of nodes
\tikzset{
  base/.style={draw, on chain, on grid, align=center, minimum height=4ex},
  proc/.style={base, rectangle, text width=14em},
  test/.style={base, diamond, aspect=2, text width=5em},
  term/.style={proc, rounded corners},
  % coord node style is used for placing corners of connecting lines
  coord/.style={coordinate, on chain, on grid, node distance=6mm and 25mm},
  % -------------------------------------------------
  % Connector line styles for different parts of the diagram
  norm/.style={->, draw},
}
% -------------------------------------------------
% place boxes
\node [proc] (p0) {1. select lens potential \\ $\phi(x,y,p)$};
\node [proc, join] (p1) {2. select parameter values $p_0$};
\node [proc, join] (p2) {3. calculate critical curves $c$\\[0.1cm] $\phi_{11}^{(0)}(p_0)\phi_{22}^{(0)}(p_0) - \left(\phi_{12}^{(0)}(p_0)\right)^2 = 0$ \\[0.1cm] $\Rightarrow c =  \left\{ x_0(p_0) \right\}$};

\node [test, join, fill=white] (t1) {fold / cusp};
\node [test,xshift=8.5em, fill=white] (t2) {$\phi_{11}^{(0)} == 0$};
\node [test, left=of t2, fill=white]  (t3)    {$\phi_{12}^{(0)} == 0$};
\node[proc, text width=8em, fill=white] (p3) {rotation matrix \\[0.1cm] $R = \left( \begin{matrix} \alpha & - \beta \\ \beta & \alpha \end{matrix}\right) $};

\node[proc, right=of p3, text width=8em, xshift=-5em, fill=white] (p4) {rotation matrix \\[0.1cm] $R = \left( \begin{matrix} 1 & 0 \\ 0 & 1 \end{matrix}\right) $};
\node[proc, right=of p4, text width=8em, xshift=-5em, fill=white] (p5) {rotation matrix \\[0.1cm] $R = \left( \begin{matrix} 0 & - 1 \\[0.1cm] 1 & 0 \end{matrix}\right) $};

\node[proc, text width=32em, yshift=-2em, xshift=-12em, fill=white] (p6) {4. rotate coordinate system \\[0.1cm] $R^\top \left( \begin{matrix} \phi_{11}^{(0)}(p_0) &  \phi_{12}^{(0)}(p_0) \\[0.1cm]  \phi_{12}^{(0)}(p_0) &  \phi_{22}^{(0)}(p_0) \end{matrix} \right)  R \Rightarrow \left( \begin{matrix} \phi_{11}^{(0)}(p_0) &  0 \\[0.1cm] 0 & 0 \end{matrix} \right) $};

\node[test,join, fill=white] (t4) {fold / cusp};
\node[proc, text width=8em,xshift=-12em, fill=white] (p7) {5. calculate \\[0.1cm]$\dfrac{\phi_{222}^{(0)}(p_0)}{\phi_{11}^{(0)}(p_0)}$};
\node[proc, text width=8em, right=of p7, xshift=7em, fill=white] (p8) {5. calculate \\[0.1cm] $\dfrac{\phi_{122}^{(0)}(p_0)}{\phi_{11}^{(0)}(p_0)}$, $\dfrac{\phi_{2222}^{(0)}(p_0)}{\phi_{11}^{(0)}(p_0)}$};

\node[proc, text width=32em, yshift=-2em, xshift=-12em, fill=white] (p9) {6. save results in database};
\node[test,join] (t5) {database complete};

\node[term, right=of t5, text width=4em, xshift=-2.5em] (p10) {END};

\begin{scope}[on background layer]
    \node[base, fit = (t1)(t2)(t3)(p3)(p4)(p5)(p6)(t4)(p7)(p8)(p9), yshift=1.4em, fill=gray!20] (loopbox) {for each $x_0(p_0)$};
\node[xshift=16.5em, yshift=-12em] (header) {$for \; each \; x_0(p_0)$};
  \end{scope}

% declare paths for test boxes
\path (t1.west) to node [near start, xshift=-1.2em, yshift=2em] {$fold$} (t3);
  \draw [->] (t1.west) -| (t3.north);
\path (t1.east) to node [near start, xshift=1.2em, yshift=1.8em] {$cusp$} (t2);
  \draw [->] (t1.east) -| (t2.north);
\path (t3.east) to node [near start, xshift=1.5em, yshift=0.5em] {$yes$} (t2);
  \draw [->] (t3.east) -- (t2.west);
\path (t3.south) to node [near start, xshift=1em] {$no$} (p3);
  \draw [->] (t3.south) -- (p3.north);
\path (t2.south west) to node [near start, yshift=-1em] {$no$} (p4);
  \draw [->] (t2.south west) -- (p4.north);
\path (t2.south east) to node [near start, yshift=-1em] {$yes$} (p5);
  \draw [->] (t2.south east) -- (p5.north);

\path (p4.south) -- (p6.north);
  \draw [->] (p4.south) -- (p6.north);
\path (p3.south) -- (p6);
\draw [->] (p3.south) -- (p6);
\path (p5.south) to node [near start, yshift=-1em] {} (p6);
  \draw [->] (p5.south) -- (p6);

\path (t4.west) to node [near start, xshift=-2em, yshift=1.8em] {$fold$} (p7);
  \draw [->] (t4.west) -| (p7.north);
\path (t4.east) to node [near start, xshift=2em, yshift=1.7em] {$cusp$} (p8);
  \draw [->] (t4.east) -| (p8.north);

\path (p7.south) -- (p9);
  \draw [->] (p7.south) -- (p9);
\path (p8.south) -- (p9);
  \draw [->] (p8.south) -- (p9);

\path (t5.east) to node [near start, xshift=1.2em, yshift=0.6em] {$yes$} (p10);
  \draw [->] (t5.east) -- (p10.west);

\node[left=of t5,xshift=4.5em] (tmp) {};
\path (t5.west) to node [near start, xshift=-3em, yshift=0.6em] {$no$} (tmp);
  \draw [->] (t5.west) -- (tmp.east) |- (p1.west);

\end{tikzpicture}
\end{center}
\caption{Flowchart to calculate ratios of derivatives at all singular points $x^{(0)}(p_0)$ for a family of lens models within a specific parameter range. A detailed description of the algorithm (including the definitions of $\alpha$ and $\beta$) can be found in Sect.~\ref{sec:implementation} adapted to the case of a singular isothermal sphere with external shear.}
\label{fig:flowchart}
\end{figure*}
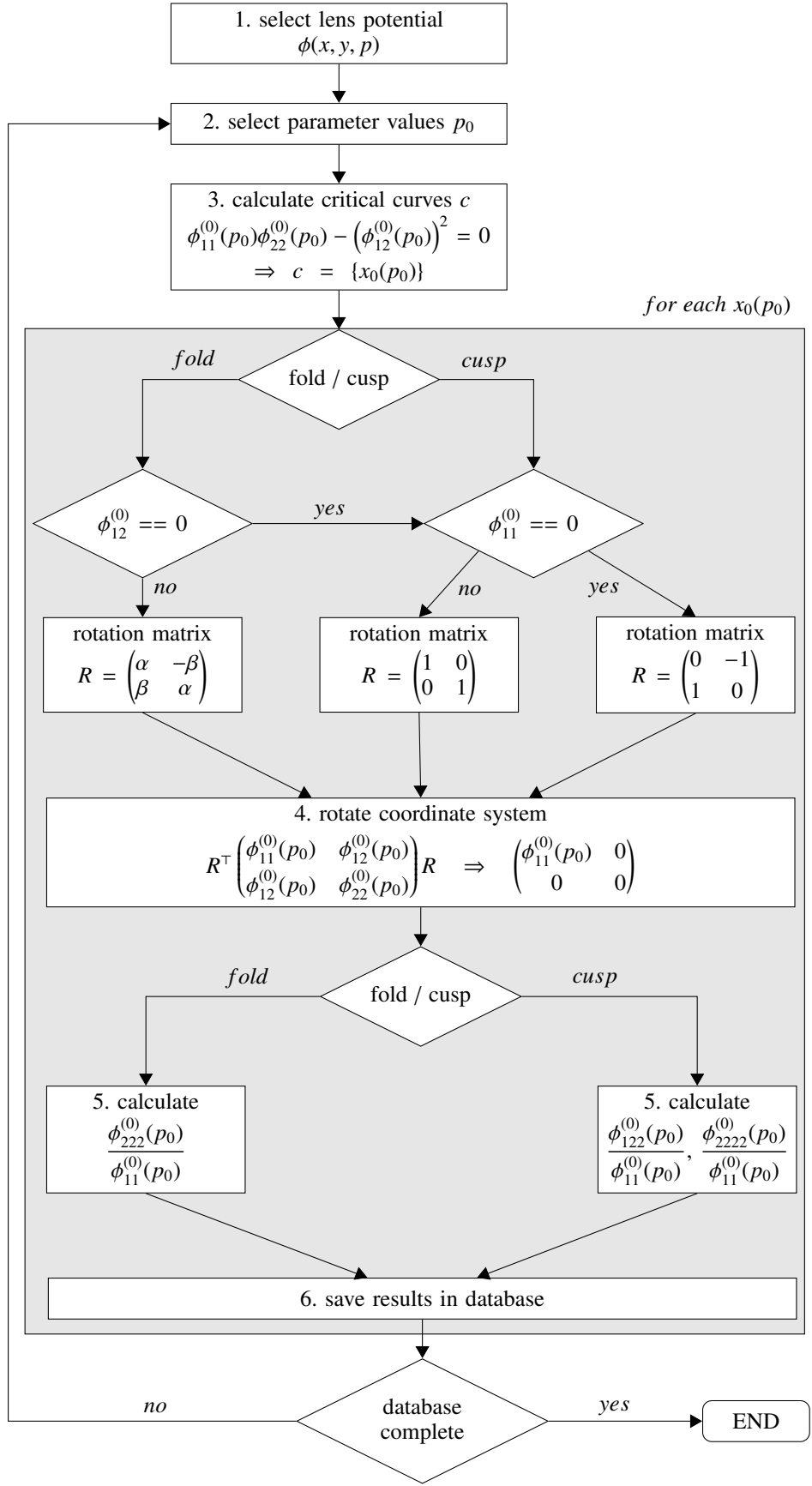

%%%%%%%%%%
\subsection{Matching observational data}
\label{sec:matching}

Having calculated the right-hand sides of Eqs.~\ref{eq:ratio_fold}, \ref{eq:ratio1_cusp}, and \ref{eq:ratio2_cusp} for a given observation together with their errors and uncertainties as described in Sect.~\ref{sec:basics}, the resulting interval of potential-derivative ratios is compared to the ratios generated according to Sect.~\ref{sec:database}. An estimate of the uncertainty in the model parameters and positions of the singular points is given by the interval of measured ratios resulting from the different combinations of observables, as detailed in Sect.~\ref{sec:errors}.

In the fold case, we uniformly sample positions $(r^{(0)}(p_0), \theta^{(0)}(p_0))$ around the critical curve, calculate the modelled ratios $\phi_{222}^{(0)}/\phi_{11}^{(0)}$ and save $p_0$, $(r^{(0)}(p_0), \theta^{(0)}(p_0))$ and the ratios as results, if the modelled ratios agree with the measured ratios within the measurement uncertainty. For the cusp case, we save $p_0$, the sampled positions $(r^{(0)}(p_0), \theta^{(0)}(p_0))$ on the critical curve and the modelled ratios $\phi_{122}^{(0)}/\phi_{11}^{(0)}$ and $\phi_{2222}^{(0)}/\phi_{11}^{(0)}$, if both of the latter are compatible with the corresponding measured ratios within the measurement uncertainty. If no match between modelled and observed ratios can be found, the entire family of lens models in the tested parameter range can be excluded from the possible model candidates for the given observation.

To illustrate the procedure outlined in Sects.~\ref{sec:database} and \ref{sec:matching}, we now apply it to the example of the SIS+E.

%%%%%%%%%%
\subsection{Parameter ranges and parameter space for SIS+E}
\label{sec:implementation_sise}

For the family of SIS+E models, we require physically motivated ranges for $a$ and $\Gamma_1$ to calculate the left-hand sides of Eqs.~\ref{eq:ratio_fold}, \ref{eq:ratio1_cusp} and \ref{eq:ratio2_cusp}. For galaxy lensing, we can estimate
\begin{equation}
a \approx \left(\dfrac{\sigma}{c}\right)^2 \approx \left(\dfrac{300 \, \mathrm{km/s}}{3 \cdot 10^5 \, \mathrm{km/s}}\right)^2 \approx 10^{-6}
\end{equation}
while for cluster lensing,
\begin{equation}
a \approx \left(\dfrac{\sigma}{c}\right)^2 \approx \left(\dfrac{1000 \, \mathrm{km/s}}{3 \cdot 10^5 \, \mathrm{km/s}}\right)^2 \approx 10^{-4} \;.
\end{equation}
In both cases, we assume small amplitudes of the external shear $\Gamma_1$ between 0 and 0.3 with sampling steps of 0.03, so that the corresponding mass distributions are approximately elliptical and all models are physically reasonable.

Thus, appropriate lens-model candidates can be expected for $a \in \left[ 1, 9 \right] \cdot 10^{-6}$ with sampling steps of $10^{-6}$ for galaxy lenses, leading to potential ratios in the ranges
\begin{align}
  -22.10 \le 10^{-5}\,&\dfrac{\phi_{122}^{(0)} }{\phi_{11}^{(0)}}\, \left( \text{rad} \right)^{-1}\le -0.47 \;, \label{eq:database_ratio1}\\
  443 \le \phantom{10^{-5}\,} &\dfrac{\phi_{222}^{(0)} }{\phi_{11}^{(0)}}\, \left( \text{rad} \right)^{-1}\le 1.01 \cdot 10^{6}\;, \label{eq:database_ratio2}\\
  0.11 \le 10^{-11}\,&\dfrac{\phi_{2222}^{(0)} }{\phi_{11}^{(0)}}\, \left( \text{rad} \right)^{-2} \le 84.18 \;. \label{eq:database_ratio3}
\end{align}
For the most massive galaxy clusters, we use $a \in \left[ 1, 9 \right] \cdot 10^{-4}$ with sampling steps of $10^{-4}$ for the same range of $\Gamma_1$ and obtain ranges for the ratios of the third-order derivatives lower by two orders of magnitude, and a ratio of the fourth-order derivative lowered by four orders of magnitude compared to Eqs.~\ref{eq:database_ratio1}, \ref{eq:database_ratio2} and \ref{eq:database_ratio3}, respectively.

Matching measured to modelled potential-derivative ratios within these parameter ranges, the resulting set of candidate models compatible with the observation can be located in the three-dimensional parameter space $(\theta^{(0)}(a, \Gamma_1), a, \Gamma_1)$, or as two-dimensional slices for fixed $\theta^{(0)}$, as shown in Sect.~\ref{sec:examples}, Fig.~\ref{fig:parameters_J2222} (3-d) and Fig.~\ref{fig:parameters_B1422} (2-d).

A more intuitive way of illustrating a model's compatibility with a given observation shows the potential-derivative ratios and the positions of the singular points where the images can be located. The plot contains the potential-derivative ratios for fixed parameter values as derived from the lens model family and the potential-derivative ratio inferred from the observation on the ordinate, and the angular position along the critical curve on the abscissa.

\begin{figure}[!h]
%\centering
\includegraphics[width=0.39\textwidth]{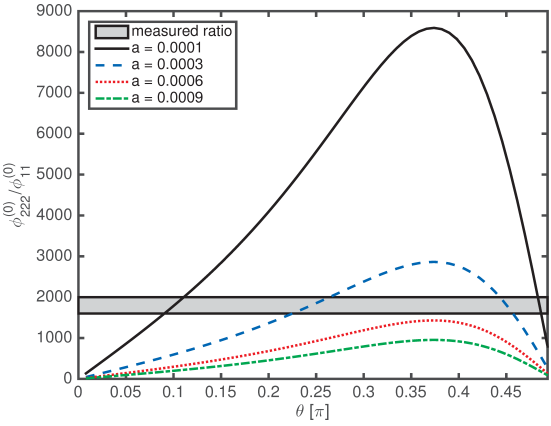}
\caption{Comparison of an (arbitrary) measured ratio of $\phi_{222}^{(0)}/\phi_{11}^{(0)}$ of an image pair at a fold singularity taking into account measuremen uncertainties and systematic errors (grey shaded area) with the lens model family of SIS+E, $a$ being chosen to cover the galaxy cluster range, as indicated by the four lines plotted for different values of $a$ within this range for an external shear of amplitude $\Gamma_1 = 0.24$.}
\label{fig:model_selection}
\end{figure}

An example is given in Fig.~\ref{fig:model_selection} assuming an (arbitrarily chosen) measured ratio of Eq.~\ref{eq:ratio_fold} near a fold caused by a galaxy cluster within the SIS+E lens model family. For different $a$ for all $x^{(0)}$ on the critical curve (black, blue, red, and green line), we observe that models with $a \ge 6 \cdot 10^{-4}$ (red and green line) can be ruled out as feasible models. In a coordinate system with its origin at the lens centre, we align the major axis of the elliptical critical curve along the $x_1$-axis. We find that possible positions on the critical curve for the two images to lie for $a = 10^{-4}$ are at angles $\theta \in \left[0.09, 0.11\right]~\pi$ or $\theta \in \left[0.48,0.49\right]~\pi$ in the first quadrant (overlap between the grey shaded area and the black line). Without time delay information, the sign of the ratio is undetermined, so that, in total, there are eight possible locations around the entire critical curve that the given observation needs to be compatible with.

%%%%%%%%%%%%%%%%%%%%%%%%%%%
\section{Examples}
\label{sec:examples}

%%%%%%%%%%
\subsection{JVAS B1422+231}
\label{sec:B1422}

\begin{figure}
  \centerline{\includegraphics[width=0.7\hsize]{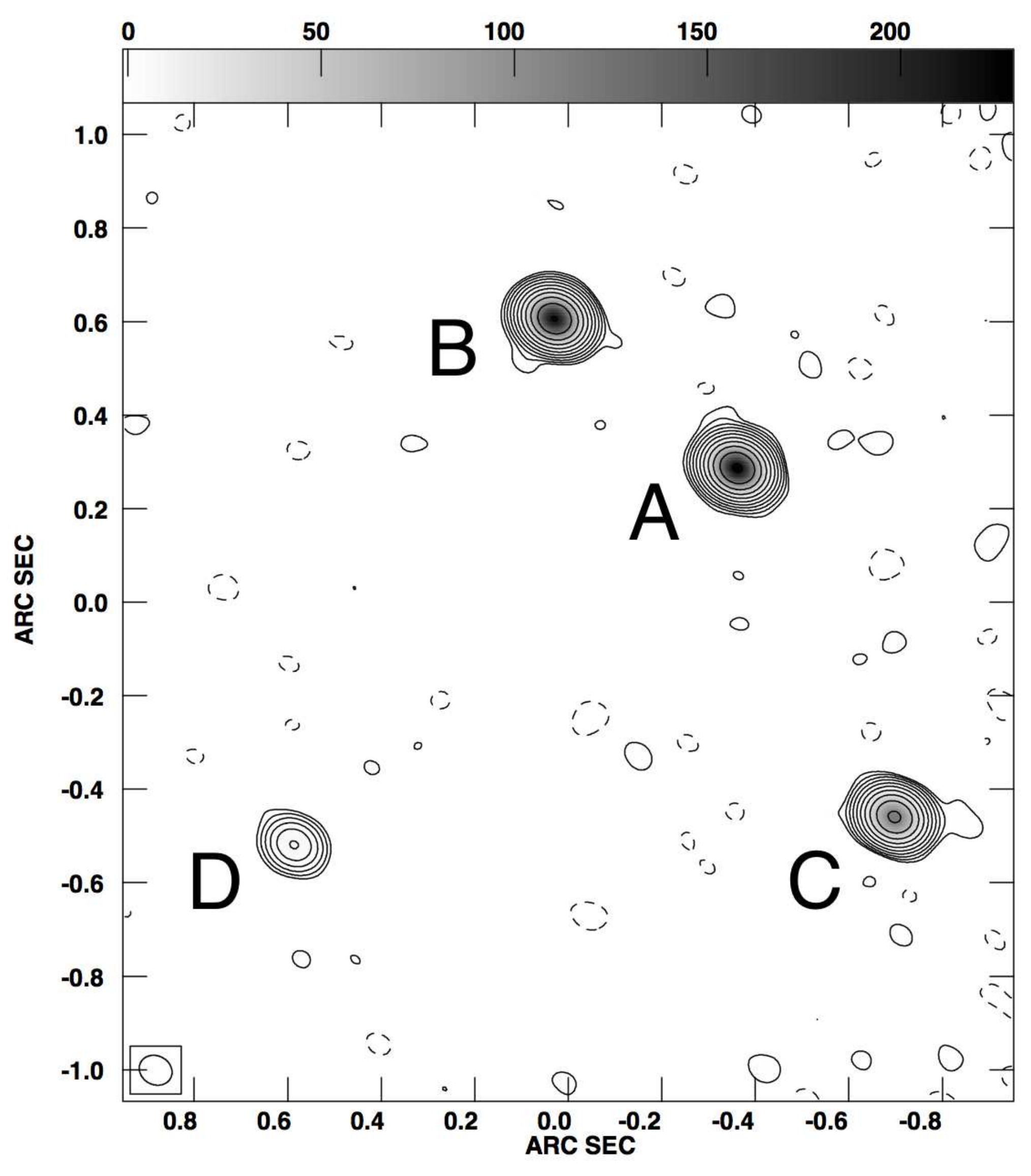}}
\caption{MERLIN map of B14122+231 at 5 GHz radio frequency, shown to define our labelling of the four gravitationally lensed images. Images $A$, $B$, and $C$ are close to a cusp singularity, $A$ being closest to the singular point. Image $D$ is on the opposite side and thus not included in our data analysis.}
\label{fig:B1422}
\end{figure}

\begin{figure}
  \centerline{\includegraphics[width=0.5\textwidth]{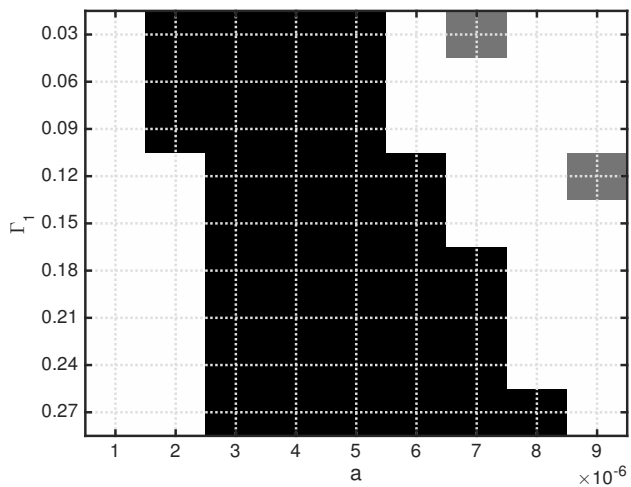}}
\caption{Plot of the parameter space $(\theta^{(0)} = \pi/2, a, \Gamma_1)$: black regions mark parameter combinations $(a, \Gamma_1)$ that are compatible with $\phi_{2222}^{(0)}/\phi_{11}^{(0)}$ from the observational data, grey regions mark areas in which only $\phi_{122}^{(0)}/\phi_{11}^{(0)}$ is compatible with observations.}
\label{fig:parameters_B1422}
\end{figure}

In \cite{bib:Wagner}, we already analysed the quadruple-image lens JVAS B1422+231, first described by \cite{bib:B1422}. Fig.~\ref{fig:B1422} shows the image labels chosen. Using the image positions measured by the \cite{bib:JVAS}, the distances in mas between the image pairs are
\begin{eqnarray}
\delta_{AB1} &= 389.25\;, &\delta_{AB2} = 319.98\;, \\
\delta_{AC1} &= 333.88\;, &\delta_{AC2} = 747.71\;, \\
\delta_{BC1} &= 723.13\;, &\delta_{BC2} = 1067.70\;.
\end{eqnarray}
The image ellipticities as determined by \cite{bib:Bradac, bib:Patnaik} are
\begin{equation}
\epsilon_A = 0.80 \pm 0.07\;, \quad \epsilon_B = 0.70 \pm 0.07\;, \quad \epsilon_C = 0.55 \pm 0.09\;.
\end{equation}
For the potential-derivative ratios using images $A$, $B$ and $C$ lying close to a cusp singularity, we obtained in \cite{bib:Wagner}
\begin{align}
  -1.622 \le 10^{-5}\,&\dfrac{\phi_{122}^{(0)} }{\phi_{11}^{(0)}}\, \left( \text{rad} \right)^{-1}\le -1.498 \;, \\
  0.12 \le 10^{-12}\,&\dfrac{\phi_{2222}^{(0)} }{\phi_{11}^{(0)}}\, \left( \text{rad} \right)^{-2} \le 1.15 \;.
\end{align}
Matching these intervals with the potential-derivative ratios of an SIS+E model, with the possible parameter ranges determined as described in \ref{sec:implementation_sise}, we immediately conclude that $a \in \left[1, 9 \right] \cdot 10^{-4}$, $\Gamma_1 \in \left[0.03, 0.3 \right]$ can be excluded. We find possible parameter combinations for $a$ and $\Gamma_1$ on the semi-major axis, i.e.\ $\theta^{(0)} = \pi/2$, as shown in Fig.~\ref{fig:parameters_B1422}\footnote{$\theta^{(0)} = 0$ can be excluded from time delay information as measured in \cite{bib:Patnaik2}.}. The black areas mark all parameter combinations for which $\phi_{2222}^{(0)}/\phi_{11}^{(0)}$ is compatible with the measurements, grey areas show compatible parameter combinations for $\phi_{122}^{(0)}/\phi_{11}^{(0)}$. Hence, we find no model that fulfils both conditions, which was to be expected beforehand from the results obtained for the singular isothermal ellipse tested in \cite{bib:Wagner}.

As a consistency check, we now perform a model selection using images $A$ and $B$ only and assume them to lie close to a fold point, such that
\begin{equation}
4.64 \le 10^{-6}\,\dfrac{\phi_{222}^{(0)} }{\phi_{11}^{(0)}}\, \left( \text{rad} \right)^{-1} \le 7.37 \;.
\label{eq:B1422_fold}
\end{equation}
Comparing Eq.~\ref{eq:B1422_fold} to the model-dependent ratios, no matching parameters are found because the highest value for this ratio for an SIS+E model is $1.01 \cdot 10^{6} \left( \text{rad} \right)^{-1}$. Yet, the database entries and measured ratios are of the same order of magnitude. Since \cite{bib:Patnaik} argue that their determination of image ellipticities may be uncertain due to poor fitting of elliptical Gaussians using \textit{JMFIT}, a more precise measurement of the image ellipticities could yield feasible model parameter combinations that are in agreement with those shown in Fig.\ref{fig:parameters_B1422}.

%%%%%%%%%%
\subsection{MACS J1149.5+2223}
\label{sec:J1149}

\begin{figure}
  \centerline{\includegraphics[width=0.35\textwidth]{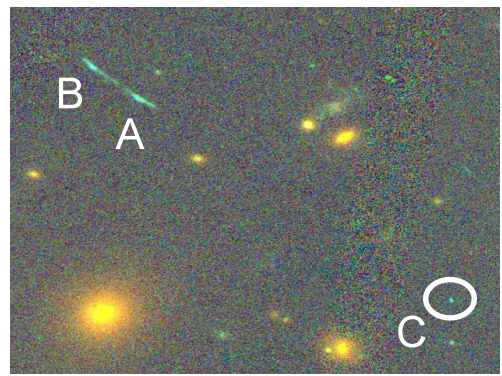}}
\caption{The three gravitationally lensed images $A$, $B$, and $C$ from \cite{bib:Wagner} used for model selection of the galaxy cluster MACS J1149.5+2223. \textit{Image credits: NASA, ESA, and M. Postman (STScI), and the CLASH collaboration.}}
\label{fig:J1149}
\end{figure}

The second example analysed in \cite{bib:Wagner} was the triple-image system in MACS J1149.5+2223 shown in Fig.~\ref{fig:J1149}. The distances between the three images in arcseconds are
\begin{equation}
\delta_{AB} = 2.42\;, \quad \delta_{AC} = 16.25\;, \quad \delta_{BC} = 18.66 \;,
\end{equation}
with a spatial resolution of 0.13 arcseconds for the image positions. The image ellipticities, determined by SExtractor, are
\begin{equation}
\epsilon_A = 0.80 \pm 0.07\;, \quad \epsilon_B = 0.70 \pm 0.07\;, \quad \epsilon_C = 0.55 \pm 0.09\;.
\end{equation}
With these observational data, we obtained singular isothermal ellipses with large ellipticities as feasible models, assuming that the image triple shown in Fig.~\ref{fig:J1149} was located near a cusp. However, as image $C$ is neither very elongated nor located in the vicinity of images $A$ and $B$, a model selection under the assumption that image $A$ and $B$ are lying close to a fold seems more reasonable. Performing this model selection, we obtain
\begin{equation}
9.12 \le 10^{-5}\,\dfrac{\phi_{222}^{(0)} }{\phi_{11}^{(0)}}\, \left( \text{rad} \right)^{-1} \le 9.15 \;.
\label{eq:phi222_J1149}
\end{equation}
We have to match this ratio with our database for massive galaxy clusters because velocity dispersions up to 1270 $\text{km s}^{-1}$ have been measured in \cite{bib:Smith}. As we derive $\sigma$ from $a$ alone, the measured velocity dispersion has to be considered as a lower bound on $\sigma(a)$ because \cite{bib:Keeton} showed that velocity dispersions in elliptical models are smaller than those in axisymmetric models when using the same normalisation constant $\rho_0$ for the mass-density profile. Finding no feasible combinations of parameters is consistent with the results of \cite{bib:Wagner}, as it is to be expected that no SIS with small to moderate external shear could fit this observation. Hence, the entire family of SIS+E for the parameter range $\Gamma_1 \in \left[0, 0.3\right]$ for any fold point along the critical curve can be ruled out.

%%%%%%%%%%
\subsection{SDSS J2222+2745}
\label{sec:J2222}

\begin{figure}
  \centerline{\includegraphics[width=0.5\textwidth]{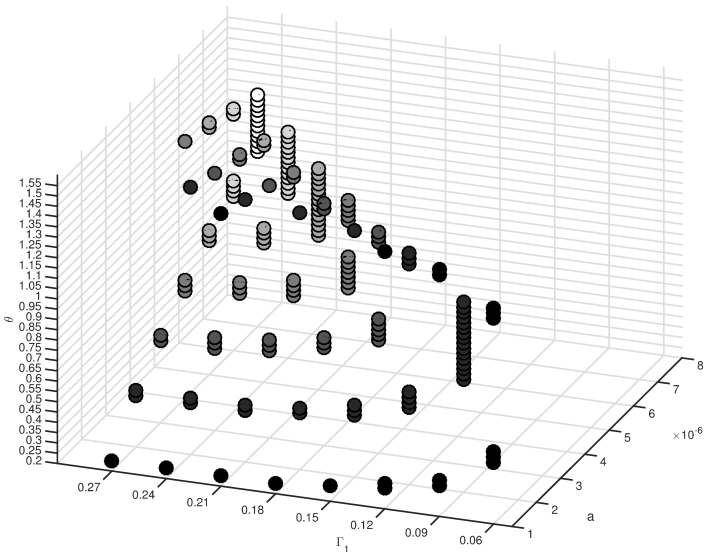}}
\caption{Plot of the parameter space $(\theta^{(0)}, a, \Gamma_1)$: circles mark parameter combinations $(\theta^{(0)}, a, \Gamma_1)$ that are compatible with the observational data with increasing grey  value for increasing $a$ for better visualisation purposes.}
\label{fig:parameters_J2222}
\end{figure}

The six images of the quasar SDSS J2222+2745, gravitationally lensed by a galaxy cluster, are the third set of images known of a quasar with maximum image separation over 10 arcseconds and the first example with five spectroscopically confirmed images, \cite{bib:Dahle1, bib:Dahle2}.
\begin{figure}
  \centerline{\includegraphics[width=0.4\textwidth]{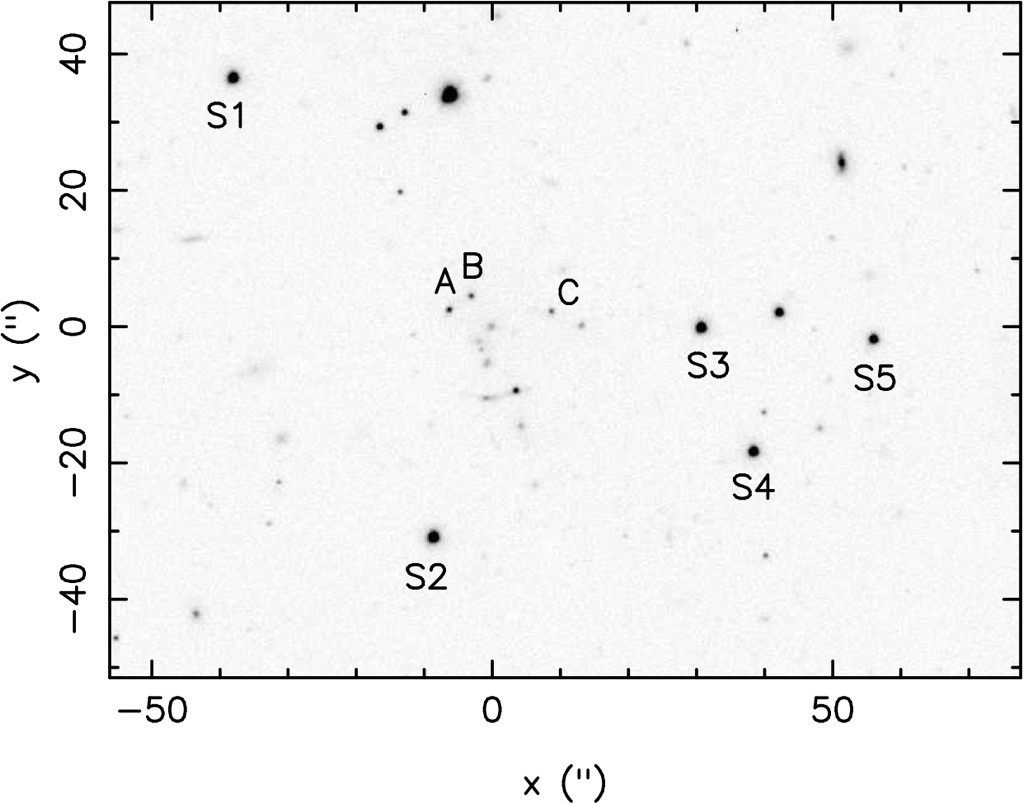}}
\caption{Three of the six gravitationally lensed images $A$, $B$, and $C$ of SDSS J2222+2745. $S1$ to $S5$ are reference stars used to calibrate the observed fluxes of the quasar images for time delay measurements. \textit{Image credits: \cite{bib:Dahle2}.}}
\label{fig:J2222}
\end{figure}
We use the images labelled $A$ and $B$ by \cite{bib:Dahle2}, as shown in Fig.~\ref{fig:J2222} taken from this work, and assume that they are located near a fold singularity. Then, we measure their relative distance to be $\delta_{AB} = 4.16 \pm 0.1$ arcseconds and determine their ellipticities by running SExtractor in its standard configuration on the $g$-band image of the field obtained from the SDSS DR12 Science Archive Server as
\begin{equation}
\epsilon_A = 0.193 \pm 0.035 \quad \epsilon_B = 0.121 \pm 0.045 \;.
\end{equation}
Evaluating the potential-derivative ratios from these measurements yields
\begin{equation}
1.26 \le 10^{-5}\,\dfrac{\phi_{222}^{(0)} }{\phi_{11}^{(0)}}\, \left( \text{rad} \right)^{-1} \le 1.466 \;.
\label{eq:phi222_J2222}
\end{equation}
Without further information from velocity dispersions, compatible SIS+E models can be found within the parameter ranges $a \in \left[1,9\right] \cdot 10^{-6}$ and $\Gamma_1 \in \left[ 0.03, 0.3\right]$. Fig.~\ref{fig:parameters_J2222} shows the set of all feasible parameter-value combinations in the three-dimensional parameter space. For better visualisation, the grey value indicating a feasible parameter value combination becomes brighter with increasing $a$. From these results, we expect velocity dispersions to fall into the range $\left[100, 300 \right] \text{km s}^{-1}$, which is too low for a galaxy cluster. Assuming that the ellipticities are overestimated, as is probable for Eq.~\ref{eq:B1422_fold} as well, possible values of $a$ lie in $\left[1,9\right] \cdot 10^{-5}$, resulting in more realistic velocity dispersions of $\left[320, 960 \right] \text{km s}^{-1}$. Another possible explanation is that the SIS+E model is inappropriate, as the cluster need not have an elliptical shape, nor an isothermal density profile, or it may contain substructures.

%%%%%%%%%%%%%%%%%%%%%%%%%%%
\section{Conclusion}
\label{sec:conclusion}

We developed an approach to connect the model-independent information provided by gravitationally-lensed images close to folds and cusps with families of lens models, such that allowed ranges of model-parameter values as well as fold and cusp locations compatible with the data can be determined, or entire model families can be ruled out. The data required for this method are the relative distances between the images and the image ellipticities. While image positions can be accurately determined, ellipticity measurements are still rare and uncertain if ellipse fitting is used, as done by \cite{bib:Patnaik}. Contributions of objects close by can also introduce a bias, e.g.\ images of SDSS J2222+2745 may contain parts of the underlying host galaxy, \cite{bib:Dahle1}. Furthermore, uncertainty estimates of the ellipticity provided by SExtractor only account for photon statistics, but can reach even up to 20\% of the measured ellipticity, so that the ellipticities are the main source of uncertainty and inaccuracy in this model selection method. Increasing the resolution of the images may reduce the uncertainty, as current ellipticity measurements are performed on strongly lensed images consisting of about 50 pixels in total. 

On a standard desktop computer, the MATLAB implementation of our model selection takes about 30 minutes to calculate approximately 5000 ratios of potential derivatives for fold and cusp positions given a lens-model family with a specific range of parameter values. Based on a database with these ratios, the actual model selection is reduced to a few seconds, matching a given range of observed ratios and retrieving the list of compatible parameter values as well as fold and cusp positions, or ruling out the model family in the given parameter range if the list remains empty.

Currently, we are working on a run-time and memory optimised Python-based version of the algorithm to calculate the ratios of potential derivatives and perform a model selection, such that users can insert their family of lensing potentials and specify the parameter ranges of interest to generate own databases of model-dependent potential-derivative ratios. The most recent version includes the calculation of the model-independent ratios of potential derivatives and for model selection the SIS+E, the singular isothermal ellipse (SIE), as described in \cite{bib:Kormann}, the Navarro-Frenk-White profile \cite{bib:NFW} and NFW profiles with elliptified radius or external shear are available.

%%%%%%%%%%%%%%%%%%%%%%%%%%%
\begin{acknowledgements}
We wish to thank Mauricio Carrasco, Felix Fabis, Robert Lilow, Matteo Maturi, Sven Meyer and S\"{o}ren Nolting for helpful discussions. We gratefully acknowledge the support by the Deutsche Forschungsgemeinschaft (DFG) WA3547/1-1.
\end{acknowledgements}

%%%%%%%%%%%%%%%%%%%%%%%%%%%
\bibliographystyle{aa}
\bibliography{model_class_selection}

\appendix

\section{Transformation of potential derivatives}
\label{sec:transformation_of_derivatives}

Denoting the potential in the coordinate system of the lens with $\tilde{\phi}$, the transformation to the coordinate system defined in Eq.~\ref{eq:coordinate_system} is given by
\begin{eqnarray}
\phi_{11}^{(0)} &=& \alpha^2 \tilde{\phi}_{11}^{(0)} + 2 \alpha \beta \tilde{\phi}_{12}^{(0)} + \beta^2 \tilde{\phi}_{22}^{(0)} \\[0.1cm]
\phi_{222}^{(0)} &=& -\beta^3 \tilde{\phi}_{111}^{(0)} + 3 \alpha \beta^2 \tilde{\phi}_{112}^{(0)} - 3 \alpha^2 \beta \tilde{\phi}_{122}^{(0)} + \alpha^3 \tilde{\phi}_{222}^{(0)} \\[0.1cm]
\phi_{122}^{(0)} &=& \alpha \beta^2 \tilde{\phi}_{111}^{(0)} + \left( \beta^3 -2 \alpha^2 \beta \right) \tilde{\phi}_{112}^{(0)}  \\
&\phantom{=}& + \left( \alpha^3 - 2 \alpha \beta^2 \right) \tilde{\phi}_{122}^{(0)} + \alpha^2 \beta \tilde{\phi}_{222}^{(0)} \\[0.1cm]
\phi_{2222}^{(0)} &=& \beta^4 \tilde{\phi}_{1111}^{(0)} -4 \alpha \beta^3 \tilde{\phi}_{1112}^{(0)} + 6 \alpha^2 \beta^2 \tilde{\phi}_{1122}^{(0)} \\
& \phantom{=}& - 4 \alpha^3 \beta \tilde{\phi}_{1222}^{(0)} + \alpha^4 \tilde{\phi}_{2222}^{(0)} \;,
\end{eqnarray}
employing the definitions of $\alpha$ and $\beta$ as given in Eq.~\ref{eq:alpha_beta}. For clarity, the dependence on $p_0$ is omitted and the translation to shift $(x^{(0)}, y^{(0)})$ to $(0,0)$ is not explicitly performed, as the latter only changes the overall position of the lens model in the coordinate system but not the values of the (ratios of) potential derivatives.

\end{document}